\documentstyle[preprint,eqsecnum,aps,floats]{revtex}
\newcommand{\drawsquare}[2]{\hbox{%
\rule{#2pt}{#1pt}\hskip-#2pt
\rule{#1pt}{#2pt}\hskip-#1pt
\rule[#1pt]{#1pt}{#2pt}}\rule[#1pt]{#2pt}{#2pt}\hskip-#2pt
\rule{#2pt}{#1pt}}
\newcommand{\Yfund}{\raisebox{-.5pt}{\drawsquare{6.5}{0.4}}}
\newcommand{\Ysymm}{\raisebox{-.5pt}{\drawsquare{6.5}{0.4}}\hskip-0.4pt%
        \raisebox{-.5pt}{\drawsquare{6.5}{0.4}}}
\newcommand{\Yasymm}{\raisebox{-3.5pt}{\drawsquare{6.5}{0.4}}\hskip-6.9pt%
        \raisebox{3pt}{\drawsquare{6.5}{0.4}}}
\newcommand{\Ythreea}{\raisebox{-3.5pt}{\drawsquare{6.5}{0.4}}\hskip-6.9pt%
        \raisebox{3pt}{\drawsquare{6.5}{0.4}}\hskip-6.9pt
        \raisebox{9.5pt}{\drawsquare{6.5}{0.4}}}

\count255=\time\divide\count255 by 60 \xdef\hourmin{\number\count255}
  \multiply\count255 by-60\advance\count255 by\time
  \xdef\hourmin{\hourmin:\ifnum\count255<10 0\fi\the\count255}
\def\equ#1{\begin{equation}#1\end{equation}}
\begin{document}
\tightenlines
\draft
\preprint{\vbox{
\hbox{UCSD/PTH 97--26}
}}

\title{Systematic Study of Theories with Quantum Modified Moduli}

\author{Benjam\'\i{}n Grinstein\cite{bg}
and Detlef R. Nolte\cite{dn}}

\address{Department of Physics,
University of California at San Diego, La Jolla, CA 92093}
\date{September 1997; Revised January, 1998}

\maketitle
\begin{abstract}
We begin the process of classifying all supersymmetric theories
with quantum modified moduli. We determine all theories based on a single
$SU$ or $Sp$ gauge group with quantum modified moduli. By flowing among
theories we have calculated the precise modifications to the algebraic
constraints that determine the moduli at the quantum level. We find a
class of theories, those with a classical constraint that is covariant but not
invariant under global symmetries, that have a singular modification
to the moduli, which consists of a new branch.
\end{abstract}



\narrowtext
\section{Introduction}

Supersymmetric gauge theories with matter fields generally have a large
degeneracy of inequivalent vacua. The space of vacua, or `moduli
space', can be readily determined at the classical level. After
quantization the problem of determining the moduli space is more
difficult because asymptotically free theories can be strongly
coupled.  Seiberg has studied the phase structure of SUSY QCD with
$N_c$ colors and $N_f$ flavors\cite{seiberg1}. It was know that the
theory has runaway vacua\cite{ADS1} for \(N_{f} < N_{c}\). Seiberg
argued that for \(N_{f} = N_{c}\) the moduli space (of vacua) is
modified by quantum effects while for \(N_{f} = N_{c}+1\) the theory
displays confinement without chiral symmetry breaking. For \(N_{f} >
N_{c}+1\) he found dual descriptions in which the magnetic dual
coupling is weak when the electric one is strong and vice versa.

Following Seiberg many have studied a number of specific SUSY gauge
theories. Intriligator and Pouliot repeated Seiberg's analysis for
$Sp(N_c)$ gauge groups with $2N_f$ flavors (matter fields in the
fundamental representation)\cite{keni1}. Pouliot\cite{pouliot} and
Trivedi and Poppitz\cite{trivedi} studied $SU(N)$ theories with an
antisymmetric tensor field and Pouliot and Strassler\cite{pouliot2}
with a symmetric tensor. Many other examples and references can be
found in reviews\cite{reviews}.

In those investigations the emphasis was on finding the behavior of a
particular theory, or class of theories.  Csaki, Schmaltz and Skiba
took a different approach\cite{CSS}. They attempted to find all
theories that display a particular effect. To this end they define
``s-confinement'' as a generalization of confinement without chiral
symmetry breaking as obtained by Seiberg for SUSY QCD with
$N_f=N_c+1$. They proceed to find all SUSY gauge theories based on a
simple gauge group that display s-confinement.

In this paper we take a similar track. We begin a study of all SUSY
gauge theories with a quantum modified moduli space. We determine all
theories based on a simple $SU(N)$ or $Sp(N)$ gauge group with a
quantum modified moduli space. We have not attempted to study
exceptional or orthogonal gauge groups.  Theories with a modified
moduli space are of interest per se. The quantum modification is
poorly understood, inferred only from consistency conditions. These
theories can be used to fabricate models of dynamical SUSY breaking
\cite{IT}.

One may describe the classical moduli space in terms
of gauge invariant composite operators. The moduli is the space of
values these operators may take, modulo algebraic constraints. At
the quantum level the description of the moduli space is still in
terms of these operators. The modification of the moduli space is to
be found in a modified algebraic constraint. It is therefore useful to
know, a priori, how many constraints one must have (given a choice of
composite operators to describe the moduli). We derive a simple
formula for the dimension of the moduli space which then gives us the
number of required constraints. 

The constraint specifying the moduli may be either invariant or
covariant under the non-anomalous global symmetries of the theory. In
the former case the quantum moduli space differs from the classical in
that the origin  has been smoothly excised. But when the
constraint is covariant the origin remains in the quantum moduli
space. In order for the 't~Hooft anomaly condition to be satisfied at
the origin, one mode must be excluded, and this can be implemented in
two distinct ways. The constraint can be used to express one mode in
terms of the rest, and therefore this mode does not contribute to the
anomaly. Alternatively one can implement the constraint with a
Lagrange multiplier in a superpotential. In this way we find that one
of the constrained modes, classically massless, becomes
massive. Integrating out this mode leaves unconstrained the rest of
the modes. The only quantum effect has been to pick which mode to
eliminate by the constraint, save for an interesting subtlety. Going
to infinity in moduli space along a particular direction we find a new
branch of moduli space. On this branch the global $U(1)_R$ symmetry is
spontaneously broken. A similar situation has been found for theories
with branches in a Coulomb phase\cite{SW1,SW2,IS}, but with the
obvious distinction that in the theories we consider there is no local
symmetry on the branch.

The methods we use are similar to those of Csaki {\it et al}.  They
used a condition on a certain sum of indices of the representations
for the particle and gauge contents.  This condition significantly
reduced the number of all possible theories. The number of theories
was further reduced by studying the flow to other theories with a
phase structure incompatible with s-confinement. The remaining theories
were checked one by one to be in the s-confining phase.

An index condition can also be used to classify theories with a
quantum modified moduli space. With the help of a generalized flow, we
not only check the phase structure of our potentially interesting
theories, but also use it to determine how the gauge invariant
operators and the constraints flow from one theory to another.  In
this way we can determine the quantum modified constraints explicitly.
One could also use our generalized flow to determine explicitly the
precise form of the constraints in the s-confining theories considered
by Csaki {\it et al}.

The paper is organized as follows. In Sect.~\ref{sec:QQMs} we classify
theories according to whether the algebraic constraint specifying the
moduli is invariant or covariant under global symmetries and discuss
the correspondingly different structure of the moduli space.  In
Sect.~\ref{index} we review the index condition for the s-confining
theories and for theories with quantum modified moduli. We also
explain there the additional conditions from the flow of the theories
and give some examples of gauge invariant operator flow. Our formula
for the number of constraints is explained in Sect.~\ref{dim-moduli}.
The methods introduced are then put to work in an explicit example in
Sect.~\ref{example}.  The results for the $SU(N)$ and $Sp(N)$ theories
are presented in Sect.~\ref{all-Qs}. We list all the theories obeying
the index condition along with their phase structure. For the theories
not yet discussed in the literature, we write down the gauge invariant
operators and the exact constraint.  We come to a conclusion in
Sect.~\ref{conclusions}.

In the appendix we list all the gauge invariant operators with their
precise index structure. This is important because there is no unique
choice of operators. Another choice will generally change the precise
form of the constraint.

It may appear that the qualitative results obtained here do not
require a precise determination of the form of the
constraints. However, care must be exercised in not choosing redundant
operators. That is, some of the operators used to described the
classical moduli space may not be independent even though it may appear
so a priori. We have found that deriving quantitatively precise
constraints guards against such errors. Moreover, we believe it will
be of general use to both model builders and field theorists to have a
complete tabulation of the precise constraints. We have undertaken the
task here.

\section{Theories with Quantum Modified Moduli}\label{sec:QQMs}
The theories with quantum modified moduli (QMM) generalize Seiberg's
SUSY QCD with \( N_{c} = N_{f} \). QMM theories are confining. The
moduli space is described by a set of composite gauge invariant
operators. A generic feature of QMM theories is that the dimension of
the vacuum is smaller than the number of independent gauge invariant
`composite' operators. Both classically and quantum mechanically the
moduli space is specified by algebraic constraints among the composite
operators. In theories with QMM the quantum and classical constraints
differ. 

Returning to our prototype, supersymmetric QCD with \( N_{c} =
N_{f}\equiv N \), we recall that the moduli is described by a matrix
valued composite $M_{ij}$ transforming as $(N,\bar N)$ under the
global symmetry group $SU(N)\times SU(N)$, and two composites $B$ and
$\tilde B$ transforming as singlets. The classical moduli is the space
of these composites, modulo the constraint \(\det(M)-B\tilde B=0\). At
the quantum level the origin is excised from the moduli space; the QMM
is described by the modified constraint \(\det(M)-B\tilde
B=\Lambda^{2N}\). Notice that the constraints remain invariant under
the global symmetry group.

In this example the origin of moduli space was taken out. This is not
generic for quantum modified theories. When the classical constraint
$F(\phi_i)=0$ for composites $\phi_i$ is covariant but not invariant
under the global symmetries, the quantum modification cannot be simply
replacing the right hand side by a non-vanishing constant. This would
break the global symmetries. Instead, the right hand side will turn
out to be of the form $\Lambda^p\phi_k$, where $\phi_k$ is a composite
with the right transformation properties under the global symmetry
group and the power $p$ is governed by dimensional analysis. 

Theories in which the constraint is covariant (c-QMM's) have different
physics than those with invariant constraints (i-QMM's). In c-QMM's
the particle corresponding to the composite $\phi_k$ that appears in
the quantum modification becomes massive. The description of the
moduli space should not include $\phi_k$. In contrast, in i-QMM's one
is free to solve the constraint for any one composite in terms of the
others.

To see this introduce a lagrange multiplier chiral superfield
$\lambda$ and use it to enforce the constraint by means of a
superpotential 
\begin{equation}
\label{eq:Wi-QMM}
W=\lambda(F(\phi_i)-\Lambda^p).
\end{equation}
for i-QMM's and
\begin{equation}
\label{eq:Wc-QMM}
W=\lambda(F(\phi_i)-\Lambda^p\phi_k).
\end{equation}
for c-QMM's.  In Eq.~(\ref{eq:Wi-QMM}) the lagrange multiplier simply
enforces the constraint $F(\phi_i)=\Lambda^p$. But in Eq.~(\ref{eq:Wc-QMM}) the
lagrange multiplier plays a dynamical physical role: it pairs up with
the composite $\phi_k$ into massive states.  Integrating out these
massive sates leaves a theory with $\phi_k$ (and $\lambda$) excluded
and a vanishing superpotential.

We believe this to be the correct realization of the constraint for
the case of theories with c-QMM. The obvious alternative is to apply
the constraint $F(\phi_i)=\Lambda^p\phi_k$ directly to the description
of the moduli. There are physical distinctions between these two
approaches, as seen in the paragraph below. Our believe in the
lagrange multiplier method can be supported by the following
argument. As discussed below, s-confining theories flow into theories
with i-QMM's. These have constraints implemented by lagrange
multipliers which can be identified with modes of the parent
s-confining theories.

Thus the field $\lambda$ must be considered a dynamical field. And this
leads to a surprising modification of the c-QMM. There is a branch
parametrized by $\lambda$ itself, with the other fields determined
from
\begin{eqnarray}
\lambda\frac{\partial F}{\partial\phi_{i}} &=& 0\qquad (i\ne k)\label{eq:dw1}\\
\lambda\left(\frac{\partial F}{\partial\phi_{ k}}-
\Lambda^p\right)&=&0\label{eq:dw2}\\
F(\phi_i)-\Lambda^p\phi_k&=&0\label{eq:dw3}
\end{eqnarray}
To solve these for arbitrary $\lambda$ generally requires that one of
the $\phi_i$ tend to infinity as some of the others approach the
origin. This may seem bizarre, but we know of no reason why such
solutions should be excluded. On this moduli subspace the $U(1)_R$
symmetry is broken. In addition, if $\phi_k$ carries any other
non-anomalous global symmetry then $\lambda$ must carry the opposite
charge and this symmetry is also broken on this branch.  The two real
scalar components of $\lambda$ can be understood as the corresponding
goldstone bosons.

\section{Index Conditions and Flows}\label{index}
\subsection{The Index Condition}\label{sub-index}
\subsubsection{The Index Condition for s-confining theories}\label{index-s}

Csaki, Schmaltz and Skiba introduced ``smooth confinement without
chiral symmetry breaking and with a non-vanishing confining
superpotential'', or ``s-confinement'' for short, as a generalization
of SUSY QCD with \(N_{f} = N_{c} + 1\). It is defined as follows. An
s-confining theory must admit a description in terms of gauge
invariant composite operators everywhere on the moduli space. The
infrared effective theory must have a smooth superpotential, ie,
polynomial in the gauge invariant operators.  The origin of the
classical moduli space must also be a vacuum of the quantum moduli
space. The definition excludes theories which admit a Coulomb phase
somewhere on the moduli space and theories which have boundaries in
the moduli space between distinct Higgs and confinement phases.
Consequently the t'Hooft anomalies should match between the short and
long distance descriptions everywhere in the moduli space, and this
was found to be true by explicit computation.

To explain the index condition for s-confining theories we need to
introduce some notation. Consider a supersymmetric theory with gauge
group $G$ and $N$ chiral matter multiplets, $Q_1,\ldots,Q_N$. In the
absence of a superpotential there are $N$ global $U(1)$ symmetries,
one for each matter field, corresponding to separate flavor number.
There is also a $U(1)$ R-symmetry. All these symmetries are broken at
the quantum level by anomalies, but one may combine the $U(1)$
R-symmetry with each of the global flavor numbers to form $N$
conserved R-symmetries, $U(1)_{R_1},\ldots,U(1)_{R_N}$ with the
following charge assignments:
\begin{displaymath}
\begin{array}{c|c|c|c|c}                                         
      &U(1)_{R_{1}} & U(1)_{R_{2}} & \cdots&U(1)_{R_{N}}   \\    \hline
Q_{1} & a_{1}       & 0            & \cdots&  0            \\
Q_{2} & 0           &  a_{2}       & \cdots&  0            \\
.     & .           &  .           & \cdots&  .            \\
.     & .           &  .           & \cdots&  .            \\
.     & .           &  .           & \cdots&  .            \\
Q_{N} & 0           &  0           & \cdots&  a_{N}   .     \\
\end{array}
\end{displaymath}
The $R$-charges $a_i$ are fixed by requiring the vanishing of the
gauge anomaly. Denoting by $\mu_G$ and $\mu_i$ the  indices of 
the adjoint and of the representation of $Q_i$, normalized to unity
for the fundamental representation, one finds:
\[a_{i} = ( \sum_{j=1}^{N} \mu_{j} - \mu_{G})/\mu_{i} .\]

Now, s-confining theories must admit a smooth superpotential. It must
carry 2 units of every one of the $R$ charges. Since only the $i$-th
field carries $R_i$ charge, it must enter the superpotential as
$Q_i^{2/a_i}$. The superpotential must be a combination of terms
of the form
\[  \Lambda^3 \prod_{i=1}^{N} (Q_{i}/\Lambda)^{2\mu_{i}/ (
\sum_{j=1}^{N}  \mu_{j} - \mu_{G})}  .\]
$\Lambda$, a dynamical mass scale, is introduced by dimensional
analysis. If there is at least one chiral superfield transforming as
the fundamental (or antifundamental) of the gauge group, which is
always the case in Csaki {\it et al}, then the smoothness of the
superpotential requires \( \sum_{j=1}^{N} \mu_{j} - \mu_{G} = 1~~{\rm
or}~~2 \). Csaki {\it et al\/} argue that, in fact, only the second solution
is available. For $Sp(N)$ theories this can be seen from Witten's
anomaly, which requires an even number of fundamentals. The index
condition for s-confinement for theories with at least one fundamental
is therefore\cite{CSS}
\[\sum_{j=1}^{N} \mu_{j} - \mu_{G} = 2 .\]

If the theory has no matter fields transforming as the fundamental
(or antifundamental) representation the index condition is relaxed: 
\( \sum_{j=1}^{N} \mu_{j} - \mu_{G} \) or 
\( (\sum_{j=1}^{N} \mu_{j} - \mu_{G})/2 \)
must be a common divisor of all the $\mu_i$.

\subsubsection{The Index Condition for Theories with Quantum Modified Moduli}

The classical constraint between the composites \(\phi_{i} \) is a non
trivial polynomial,
\[  \sum_{n=1}^{m} (\prod_{i=1}^{k_{n}} \phi_{i})_{n} = 0 .\]
The quantum modification generically is of the form
\equ{\label{Qconst}
\sum_{n=1}^{m} (\prod_{i=1}^{k_{n}} \phi_{i} )_{n} 
= \prod_{i} \phi_{i} \Lambda^{p} .}
Notice that we have allowed for a product of composites on the right
hand side. In all the cases we study we find, however, at most one
composite on the right hand side.  The exact form of the left hand
side, \(\sum_{n=1}^{m} (\prod_{i=1}^{k_{n}} \phi_{i})_{n} \), is
determined by the classical limit.

The index condition now follows from requiring that the constraint be
covariant under global $U(1)$ R-symmetries. As in our review of
s-confining theories we introduce an anomaly free $U(1)$ R-symmetry
for each chiral superfield. Because the left and  right sides of the
constraint in Eq.~(\ref{Qconst}) have different number of composites,
at least one of the R-charges must vanish. For this
we must have the index condition\cite{CSS}
\equ{\label{index-const}\sum_{i=1}^{n} \mu_{i} - \mu_{G} = 0 .}

In an alternative derivation of the index condition we adopt the
point of view that \(\Lambda^{b_{0}}\) is a background chiral
superfield. Now consider the $R$ symmetry with all the $R$ charges of
the chiral superfields set to vanish. The assigned  R-charge of
\(\Lambda^{b_{0}}\) is given by the anomaly,
\[Q_{R}(\Lambda^{b_{0}})=\sum_{i=1}^{n} \mu_{i} - \mu_{G}  .\]
The left side of our constraint, however, has an R-charge of
zero. Therefore, \(\Lambda^{b_{0}}\) has an R-charge of zero and we
have again Eq.~(\ref{index-const}).

To find all QMM theories one must begin by classifying all theories
that satisfy Eq.~(\ref{index-const}). Since the fundamental
representation has $\mu_{\rm fund}=1$, adding a pair of chiral
superfields, one in the fundamental and one in the antifundamental
representations, to a QMM theory gives a theory with \(\sum_{i=1}^{n}
\mu_{i} - \mu_{G} = 2 \). These are candidates for s-confinement and
were classified by Csaki {\it et al}. Therefore all theories
satisfying the index condition (\ref{index-const}) can be obtained from
the list of s-confinement candidates of Csaki {\it et al} by removing
a fundamental and an antifundamental. Clearly removing a pair
ensures that all the gauge anomalies remain
absent. Section~\ref{all-Qs} contains tables listing the complete set
of QMM candidates based on $SU$ and $Sp$ gauge groups.

\subsection{The Flow}
\subsubsection{The Flow of the Theories}

The index condition gives only a necessary condition. To find out if a
candidate theory actually has a QMM, one must make some other
investigations.  As the next step to sort out all QMM theories we
consider points in the classical moduli space where the gauge group of
our candidate theory is broken. The gauge fields which correspond to
the broken generators acquire a mass proportional to the vacuum
expectation value of the Higgs field. These massive gauge superfields
pair up with chiral superfields which become massive through the Higgs
mechanism as well. Together they form a massive supermultiplet. We
integrate out these heavy degrees of freedom. The new theory, which is
an effective theory of the original `UV' theory, should be in a phase
consistent with the UV theory being in a quantum modified phase.  We
refer to this as `the flow' of the UV theory to an effective
theory. If the theory flows to a theory in a Coulomb phase we say that
the theory has a Coulomb branch, not a QMM. By studying the flow we
can, therefore, rule out quite a few theories which fulfill the index
condition.

It is useful to tabulate the manner in which theories may flow.  Below
we list the gauge groups together with their particle content. The
latter is contained in square brackets and is represented by the Young
tableaux of the corresponding representation, with a possible
multiplier when there are more than one field for that
representation. We don't list any gauge singlets that may remain in
the effective theory. These are not all the possible flow
diagrams. They were, however, sufficient for our classification work.

\begin{eqnarray}
\label{flow1}
SU(N) [\; N (\Yfund + \overline{\Yfund}) \;] \longrightarrow SU(N-1) [\; (N-1)
(\Yfund + \overline{\Yfund}) \;]
\end{eqnarray}
\begin{eqnarray}
\label{flow2}
SU(N) [\;\Yasymm + (N-1)\, \overline{\Yfund} + 3\, \Yfund \;] &
\longrightarrow & SU(N-1) [\; \Yasymm + (N-2)\, \overline{\Yfund} +
3\, \Yfund \;]\nonumber 
\\ \downarrow\qquad\qquad & & \\ 
Sp(N) [\; (N+2)\, \Yfund \;] &
\longrightarrow & Sp(N-2) [\; (N)\, \Yfund \;]\nonumber
\end{eqnarray}
\begin{eqnarray}
\label{flow3}
SU(N) [\;\Yasymm + \overline{\Yasymm} + 2 (\Yfund +
\overline{\Yfund}) \;] 
& \longrightarrow & 
SU(N-1) [\;\Yasymm +
\overline{\Yasymm} + 2 (\Yfund + \overline{\Yfund}) \;] \nonumber\\ 
\downarrow\qquad\qquad & &
\\ Sp(2N)[\; \Yasymm +4\, \Yfund\;] &&\nonumber
\end{eqnarray}
\begin{eqnarray}
\label{flow4}
SU(N) [\; {\rm Adj} \;] \longrightarrow \mbox{Coulomb branch}
\end{eqnarray}
\begin{eqnarray}
\label{flow5}
SU(6) [\; \Ythreea + 3 (\Yfund + \overline{\Yfund}) \;] &
\longrightarrow & SU(3) \times SU(3) [\; 3( (1,\overline{\Yfund}) +
(\overline{\Yfund} ,1) +(1,\Yfund) + (\Yfund{},1) ) \;] \nonumber\\ 
\downarrow\qquad\qquad & &\nonumber \\ 
SU(5) [\; 2\, \Yasymm + 1\, \Yfund + 3\, \overline{\Yfund} \;] &
\longrightarrow & Sp(4) [\; (\Yasymm + 4 \Yfund ) \;] \\ 
\downarrow\qquad\qquad & &\nonumber \\ 
SU(4) [\; 2\, \Yasymm + 2 (\Yfund + \overline{\Yfund}) \;] &&
(\mbox{This is special case of } SU(N) [\; \Yasymm +
\overline{\Yasymm} + 2 (\Yfund + \overline{\Yfund}) \;].)\nonumber
\end{eqnarray}  
\begin{eqnarray}
\label{flow6}
SU(4) [\; 3\, \Yasymm + 1 (\Yfund + \overline{\Yfund}) \;] &
\longrightarrow & Sp(4) [\; 2\, \Yasymm +2\, \Yfund \;]
\longrightarrow (SU(2) \times SU(2)) [\; (\Yfund,1) + (1,\Yfund) +
(\Yfund,\Yfund) \;]\nonumber \\ 
\downarrow\qquad\qquad & & \\ 
SU(3) [\; 3 (\Yfund +
\overline{\Yfund}) \;] & &\nonumber
\end{eqnarray}
\begin{eqnarray}
\label{flow7}
 SU(4) [\; 4\, \Yasymm \;] \longrightarrow Sp(4) [\; 3\, \Yasymm \;]
 \longrightarrow \mbox{Coulomb branch}
\end{eqnarray}
\begin{eqnarray}
\label{flow8}
SU(5) [\; 2\, \Yasymm + \overline{\Yasymm} + 1\, \overline{\Yfund} \;]
\longrightarrow Sp(4) [\; 2\, \Yasymm +2\, \Yfund \;] \longrightarrow
(SU(2) \times SU(2)) [\; (\Yfund,1) + (1,\Yfund) + (\Yfund,\Yfund) \;]
\end{eqnarray}
\begin{eqnarray}
\label{flow9}
SU(6) [\; 2\, \Ythreea \;] \longrightarrow SU(3) \times SU(3)[\;
((\Yfund,\overline{\Yfund}) + (\overline{\Yfund},\Yfund)) \;] \longrightarrow
\mbox{Coulomb branch}
\end{eqnarray}
\begin{eqnarray}
\label{flow10}
SU(7) [\; \Ythreea + 4\, \overline{\Yfund} + 2\, \Yfund \;] &
\longrightarrow & SU(3) \times SU(3) [\; (3 (1,\overline{\Yfund}) +
(\overline{\Yfund},1)) + (\Yfund,\Yfund) \;]\nonumber \\ 
\downarrow\qquad\qquad & & \nonumber\\
SU(6) [\; \Ythreea + \Yasymm + 2\, \overline{\Yfund} \;] &
\longrightarrow & SU(3) \times SU(3) [\; (3 (1,\overline{\Yfund}) +
(\overline{\Yfund},1)) + (\Yfund,\Yfund) \;]\nonumber \\ 
\downarrow\qquad\qquad & & \\
Sp(6) [\; \Ythreea +3\,\Yfund \;] & \longrightarrow & Sp(4) [\; 2\,
\Yasymm +2\, \Yfund \;] \longrightarrow (SU(2) \times SU(2))[\;
(\Yfund,1) + (1,\Yfund) + (\Yfund,\Yfund) \;]\nonumber \\ 
\downarrow\qquad\qquad & &\nonumber \\
SU(3) [\; 3 (\Yfund + \overline{\Yfund}) \;] & &\nonumber
\end{eqnarray}
\begin{eqnarray}
\label{flow11}
Sp(2N) [\; \Ysymm ={\rm Adj} \;] \longrightarrow \mbox{Coulomb branch}.
\end{eqnarray}

\subsubsection{Flow of Operators}

The flow is useful in determining the quantum modified constraints
precisely. For example given a classical constraint, one can
immediately write a putative quantum modified constraints as in
Eqs.~(\ref{eq:Wi-QMM}) or~(\ref{eq:Wc-QMM}). The precise coefficient is
then determined by flowing to a theory with exactly known
constraint, such as SUSY QCD with $N_f=N_c$.

This works because the flow maps not just theories but also specific
operators between the UV and effective theories.  This is very useful
in the determination of the classical constraints too.  Given a
classical constraint in a UV theory one can generally determine the
constraint in any of its effective theories by following the flow. It
is not so obvious that one may infer the constraint of a UV theory if
the constraints of its effective theories are known. In fact, in
practice one finds that for many cases one needs only the constraints
of one of the effective theories. We found this reverse flow procedure
of central importance in our investigations of the more complicated
theories.

For example, one can start from the known theory \(SU(3) [\;3 (\Yfund
+ \overline{\Yfund})\;]\) and map the gauge invariant operators up to
\(SU(4) [\;3\, \Yasymm + 1 (\Yfund + \overline{\Yfund})\;] \). Then mapping
down to \(Sp(4) [\;2\, \Yasymm +2\, \Yfund\;]\) is possible. One can
then determine how the operators are mapped from one theory to the
next. For example, 
\[ SU(4) [Q\bar Q] \rightarrow Sp(4) [Q_{1}Q_{2}] \]
where $Q\bar Q$ and $Q_{1}Q_{2}$ represent composites of the $SU(4)$
and $Sp(4)$ theories, respectively (if the notation is not
self-evident, it will be clarified in Section~\ref{all-Qs}).  Thus one
can find all the constraints of these $SU(4)$ and $Sp(4)$
theories. The constraints are determined explicitly, that is, all the
numerical coefficients are fixed.  One can obtain all the results for
the remaining theories with similar sequences of reverse and forward
flows.

Mapping the gauge invariant operators from a smaller theory to a
bigger theory may be problematic. The Higgs mechanism may map some of
the gauge invariant operators to singlets in the effective theory or may
even render some of the gauge invariant operators in the UV theory
equal to zero. This implies that we would break our theory to a place
of the moduli space which has a dimension strictly smaller by more
than one. The zero operators or the singlet operators would not show
up and all the terms in which they do show up would not be in the
constraint if we map the constraint from the effective theory to the
UV theory. This happened only twice in our analysis. When this
happened, however, we could always determine  the constraint by
requiring covariance under global symmetries.

Other useful examples of operator mapping are presented in the
sequences below, in which $Q$ and $\bar Q$ always stand for a
fundamental and anti-fundamental ($\Yfund$ and $\overline{\Yfund}$), and
$A$ and $B$ for a two and three index antisymmetric tensors
(\,$\Yasymm$ and $\Ythreea$\,), respectively. The composite operators
are denoted by their components, and a subscript ``anti'' is included
when only the antisymmetric part is included (the precise
description of the operators can be found in the appendix). The
mappings are
\[ SU(7) [B^{3}{\bar Q}^{3}Q_{anti} ] 
\rightarrow SU(6) [(B^{2}A{\bar Q}^{2})_{anti} ] 
\rightarrow Sp(6)
[BQ^{3}] \rightarrow Sp(4) [Q_{1}Q_{2}] \]
for the theory flow in (\ref{flow10}), and
\[ SU(5) [A^{2}{\bar A}{\bar Q}^{2}] \rightarrow Sp(4) [Q_{1}Q_{2}] \]
for the flow in (\ref{flow8}).
%

\section{Dimension of the moduli space}
\label{dim-moduli}
We derive formulas for the dimension of the classical moduli
space. These formulas give a relation between the number of gauge
invariant operators and the number of constraints.

Because the SUSY lagrangian is invariant under the complexified gauge
group $G_{c}$, the moduli space $M_{0}$ is equal to

\[ M_{0} = F \| G_{c} .\]

F is the space of all constant field configurations if there is no
superpotential. It is the space of all extrema of the superpotential
if there is a superpotential. The equivalence relation between two
elements $\Phi$ and $\Phi_{0}$ of the same $G_{c}$ orbit is of the
generalized form $\lim g_{i} \Phi = \Phi_{0}$ with $g_{i} \in G_{c}$.

$F \| G_{c}$ can be described as an algebraic variety of all
gauge invariant holomorphic polynomials\cite{LT}. Therefore the dimension of
the vacuum is
\begin{equation}
\label{dimvacops} 
\dim {\rm vacuum} =N_{\rm Ops} - N_{\rm Con} 
\end{equation} 
where $N_{\rm Ops}$ and $N_{\rm Con}$ are the number of independent
gauge invariant operators and constraints, respectively.  But there is
a natural map $\pi$ between F and $M_{0}$. This map induces a map
between the tangent spaces of F at the generic point $\phi \in F$ and
the tangent space at the point $\pi(\phi)$.  This map is a surjective
homomorphism if $\phi$ is a point on the moduli space which breaks the
gauge group totally. The kernel is $G_{c}\phi$\cite{GA}. That $\phi$
is on the moduli space implies of course that the D-flat condition is
fulfilled.  It follows directly that:

\begin{equation}
\label{dimvac}
\dim {\rm vacuum} = \dim F - \dim G_c .
\end{equation} 
s-confining theories and quantum modified theories have no
superpotential and they have always points on the moduli space where
the gauge group is completely broken.

The two formulas for the dimension of the moduli space allow us to
calculate easily the difference between the number of gauge invariant
operators and constraints. 
As an example consider $ SU(5)$ with $(2\Yasymm + \Yfund +
3\overline{\Yfund})$ (example 4.1.3 in Ref.~\cite{CSS}). There are 18
gauge invariant operators and the dimension of the moduli space, as
given by Eq.~\ref{dimvac} is 16, so that there must be {\it two}
constraints. 
The constraints are  easily obtained by integrating out
 $(Q_{2}\overline{Q}_{4})$ and
$(A^2Q_{1}Q_{2}\overline{Q}_{4})$ from the superpotential of the
corresponding s-confining theory, $ SU(5)$ with $(2\Yasymm + 2\Yfund +
4\overline{\Yfund})$ (here we are using the notation of
Ref.~\cite{CSS}).  Alternatively one can use the operator flow between
$ SU(5)$ with $ (2\Yasymm + \Yfund + 3\overline{\Yfund})$ and $ SU(4)$
with $ (2\Yasymm + 2\Yfund + 2\overline{\Yfund})$ to map the two
constraints for the SU(4) theory to the constraints of the SU(5)
theory.  One obtains two constraints, one is quantum modified and the
other is not. The corresponding superpotential is: 
\[ W = \lambda
[(A^3\overline{Q})^2 (Q\overline{Q}) +
(A^3\overline{Q})(A^2Q)(A\overline{Q}^2) - \Lambda^{10}] + \mu
[(A^3\overline{Q}_{i})^{a}(A\overline{Q}^2_{jk})^{b} \epsilon^{ijk}
\epsilon_{ab}].\]

\section{An Example}\label{example}
Before giving our results we present an example in which we apply all
the tools presented above. Consider\cite{CSS} an $SU(4)$ gauge theory with 3
antisymmetric tensors $A_{\alpha\beta}^i$, a fundamental $Q_\alpha$
and an antifundamental $\bar Q^\alpha$. Since $\mu_G=8$, $\mu_A=2$ and
$\mu_Q=\mu_{\bar Q}=1$ we see that the index condition,
\(\sum_{i=1}^{n} \mu_{i} - \mu_{G} = 0\), is satisfied. Adding an
additional $Q$, $\bar Q$, gives a theory with 
\(\sum_{i=1}^{n} \mu_{i} - \mu_{G} = 2\) which is not s-confining.

According to Eq.~(\ref{dimvac}) the dimension of the classical moduli
space is  $3\times6+2\times4-15=11$. To determine the number of
constraints we need a choice of composites. Consider the obvious
choice
\begin{eqnarray}
(AA_{\rm sym})^{ij}  & =& A_{\alpha \beta}^{i} A_{\gamma \delta}^{j} 
\epsilon^{\alpha \beta \gamma \delta}\\
(AAQ{\bar Q})^{ij}& =& A_{\alpha \beta}^{i} A_{\gamma \delta}^{j} 
Q_{\eta}{\bar Q}^{\alpha} \epsilon^{\beta \gamma \delta \eta}
\end{eqnarray}
It would seem that these 15 operators are sufficient to characterize
the 11 dimensional moduli space if four constraints are
imposed. However subspaces of the moduli characterized by $A=0$ with
arbitrary $Q=\bar Q^\dagger$ are not properly parametrized by these
composites. We see that we need in addition
\begin{equation}
(Q{\bar Q})  = Q_{\alpha}{\bar Q}^{\alpha} 
\end{equation}
This set of operators is not independent. One can verify that the part
of $(AAQ{\bar Q})$ symmetric under $i\leftrightarrow j$ is proportional
to $(Q{\bar Q})(AA_{\rm sym})$. We do not consider this a
constraint, for the relation involves $(AAQ{\bar Q}_{\rm sym})$
linearly: one should simply exclude this operator. 

What operators might we need, in addition to $(Q{\bar Q})$, 
$(AA_{\rm sym})$ and
$(AAQ{\bar Q}_{\rm anti})$, to describe the moduli? To answer this we
flow to $SU(3)$, along directions of non-vanishing $Q=\bar Q^\dagger$,
as in Eq.~(\ref{flow6}). This theory is the familiar example analyzed by
Seiberg and has a classical constraint \(\det(M)-B\tilde B=0\)
involving baryons. However none of the operators above flow to these
baryons. To remedy this we include in our list

\begin{eqnarray*}
(AAA{\bar Q}{\bar Q}) &=&1/6 ({\bar Q}^{\alpha} A_{\alpha \beta}^{i} 
{\bar Q}^{\gamma} A_{\gamma \delta}^{j} A_{\eta \iota}^{k}  
\epsilon^{\beta \delta \eta \iota} \epsilon_{i j k}) \\
(AAAQQ) &=& 1/6 (A_{\alpha \beta}^{i} A_{\gamma \delta}^{j} A_{\eta \iota}^{k} 
Q_{\kappa} Q_{\lambda} \epsilon^{\kappa \delta \eta \iota} 
\epsilon^{\alpha \beta \gamma\lambda }\epsilon_{i j k} )
\end{eqnarray*}
which flow to $B$ and $\tilde B$.

The set of operators
$(Q{\bar Q})$, $(AA_{\rm sym})$, $(AAQ{\bar Q}_{\rm anti})$, $(AAA{\bar
Q}{\bar Q})$ and $ (AAAQQ)$  is what we list in 
Sect.~\ref{su4}. With $N_{\rm Ops}=12$ Eq.~(\ref{dimvacops}) implies we
need one constraint. The constraint must flow to \(\det(M)-B\tilde
B=0\) in $SU(3)$. Now, $(Q{\bar Q})(AA_{\rm sym})+(AAQ{\bar Q}_{\rm
anti})$ flows to $M$ and $(AAA{\bar Q}{\bar Q})$ and $( AAAQQ)$ flow to $B$
and $\tilde B$. It follows that the classical constraint must be of
the form \( \det[(Q{\bar Q})(AA_{\rm sym})+(AAQ{\bar Q}_{\rm anti})]-
(AAA{\bar Q}{\bar Q})( AAAQQ)=0\), or by expanding the determinant and
keeping track of numerical constants 
\begin{equation}
\label{su4-class-const}
1/6  (AA_{sym})^3 (Q{\bar Q})^2 + 4 (AA_{sym})(AAQ{\bar Q}_{anti})^2 + 
64 (AAA{\bar Q}{\bar Q}) (AAAQQ) =0 
\end{equation}
where
\begin{eqnarray*}
(AA_{sym})^3  & = &  (AA_{sym})^{ij} (AA_{sym})^{kl} (AA_{sym})^{mn} \epsilon_{ikm} 
\epsilon_{jln} \\
(AAQ{\bar Q}_{anti})^2 (AA_{sym})  & = &  (AAQ{\bar Q})^{[ij]} (AAQ{\bar
Q})^{[kl]}  (AA)^{mn} \epsilon_{i j m} \epsilon_{k l n}.
\end{eqnarray*}

It is now a simple exercise to verify this constraint (with the help
of symbolic manipulator programs).

To explore the quantum moduli we note, as above, that the 't~Hooft
anomaly matching conditions are satisfied everywhere except at the
origin which must therefore be excluded by modifying the classical
constraint.  The theory has a non-anomalous global $U(1)$ symmetry
under which the fields $A$, $Q$ and $\bar Q$ transform with charges 1,
$-3$ and $-3$, respectively. The left hand side of the constraint in
Eq.~(\ref{su4-class-const}) transforms non-trivially, with charge
$-6$. This is an example of a c-QMM. The composite $(Q\bar Q)$ has charge
$-6$. It is straightforward to check that the constraint
\begin{equation}
\label{su4-quant-const}
1/6  (AA_{sym})^3 (Q{\bar Q})^2 + 4 (AA_{sym})(AAQ{\bar Q}_{anti})^2 + 
64 (AAA{\bar Q}{\bar Q}) (AAAQQ) =\Lambda^{8}(Q{\bar Q})
\end{equation}
flows to the corresponding $SU(3)$ constraint.

This c-QMM constraint does not exclude the origin of moduli space
where the 'tHooft anomaly condition is not satisfied. However, if the
constraint is implemented by a lagrange multiplier, $\lambda$, via a
superpotential
\[
W=\lambda[
1/6  (AA_{sym})^3 (Q{\bar Q})^2 + 4 (AA_{sym})(AAQ{\bar Q}_{anti})^2 + 
64 (AAA{\bar Q}{\bar Q}) (AAAQQ) - \Lambda^{8}(Q{\bar Q})],
\]
and  $\lambda$ is interpreted as a dynamical field, both
$\lambda$ and $Q{\bar Q}$ become massive. This removes one composite
from the spectrum and leaves the others unconstrained, and the
't~Hooft anomaly matching conditions are satisfied.

This superpotential exhibits a new, purely quantum mechanical, branch
of the moduli space. Consider directions on the moduli given by the
scalings $(AA_{sym})\sim\epsilon^{-1}$, $(Q{\bar Q})\sim\epsilon^3$,
$(AAQ{\bar Q}_{anti})\sim\epsilon^{1+x}$ (any $x>0$) and $(AAA{\bar
Q}{\bar Q})= (AAAQQ)=0$. These are in the moduli only if $\lambda=0$,
but in the limit $\epsilon\to0$ the moduli includes the branch
$\lambda\neq0$. Since $\lambda$ carries 2 units of $U(1)_R$, the
symmetry is spontaneously broken on this branch. Although
$(AA_{sym})\to\infty$, there remains at least an unbroken $SU(2)$
gauge group, which is strongly coupled in the neighborhood of this
branch. This suggests the interpretation of $\lambda$ as a
glueball superfield, and $\lambda\neq0$ as gaugino condensation.

\section{All Quantum Modified Theories}\label{all-Qs}
This section contains our results.  In tables~\ref{SUtable}
and~\ref{SPtable} we list all gauge and Witten anomaly free theories
that satisfy the index constraint~(\ref{index-const}) for $SU$ and $Sp$
gauge groups, respectively. We give the gauge
group in the first column, the matter content in the second and state
whether the theory has a QMM or a Coulomb branch in the last column.

For all theories which are derived from an s-confining theory by
taking out a fundamental and antifundamental, the superpotential is
easily determined by integrating out a fundamental and an
antifundamental. This was done by Csaki {\it et al} and we do no
reproduce their results here.\footnote{However,
we do  not agree with some of their results. For details see section
\ref{dim-moduli}. }
For the rest of the theories, those which only follow from theories
with \(\sum_{j=1}^{N} \mu_{j} - \mu_{G} = 2 \) that are not
s-confining, we use the flow to determine the classical constraint.

Next, in separate sub-sections, we give the precise results for those
theories which cannot be obtained from an s-confining
theory by integrating out a fundamental-antifundamental pair. 
In each case we give a table. The upper part of each table lists the
chiral superfields, the representation they belong to under the gauge
group and finally their global symmetry properties. The second part of
the table shows the analogous information for the composite
operators. The composite operators are labeled by their component
fields. There is often more than one way to construct an invariant
operator from the given component fields. The precise construction
used is specified in the appendix. In some tables there is a third
part which introduces shorthand notation convenient for giving the
constraint.  The explicit constraint is then given.

\subsection{The Quantum Modified $SU(N)$ Theories}

\begin{table}[h]
\begin{center}
\begin{tabular}{|l|l|l|} \hline
$SU(N)$   &  $ N (\Yfund + \overline{\Yfund})$   & i-quantum modified    \\
$SU(N)$   &  $\Yasymm + (N-1)\, \overline{\Yfund} + 3\, \Yfund $ 
                                              & i-quantum modified    \\
$SU(N)$ & $\Yasymm + \overline{\Yasymm} + 2 (\Yfund +
\overline{\Yfund})$     		& i-quantum modified    \\
$SU(N)$ & $Adj  $ & Coulomb branch \\ \hline
$SU(4)$ & $3\, \Yasymm + 1 (\Yfund + \overline{\Yfund})$ & c-quantum modified  \\
$SU(4)$ & $ 4\, \Yasymm $ & Coulomb branch \\ \hline
$SU(5)$ & $ 2\, \Yasymm + 1\, \Yfund + 3\, \overline{\Yfund}$ 	
					& i-quantum modified  \\
$SU(5)$ & $2\, \Yasymm + \overline{\Yasymm} + 1\, 
	\overline{\Yfund}$ & c-quantum modified  \\ \hline
$SU(6)$ & $2\, \Yasymm + 4\, \overline{\Yfund}$ & i-quantum modified  \\
$SU(6)$ & $\Ythreea + 3 (\Yfund + \overline{\Yfund})$ & i-quantum modified  \\
$SU(6)$ & $\Ythreea + \Yasymm +  2\, \overline{\Yfund}$ 
& c-quantum modified  \\
$SU(6)$ & $ 2\, \Ythreea $ & Coulomb branch  \\  \hline
$SU(7)$ & $ \Ythreea + 3\, \overline{\Yfund} + 1\, 
		\Yfund$  & c-quantum modified  \\  \hline
\end{tabular}
\end{center}
\caption{These are all $SU$ theories satisfying $\sum_j \mu_j -\mu_G =
0$ and free of gauge anomalies.  We list the gauge group and
the field content of the theories in the first and second column. In
the third column, we indicate whether the theory has a quantum
modified moduli space or a Coulomb branch. The prefix ``i'' indicates 
an invariant quantum modification and the prefix ``c'' a covariant 
quantum modification.}
\label{SUtable}
\end{table}

\subsubsection{$SU(4)$ with $ 3 \protect\Yasymm +(\protect\Yfund +
               \overline{\protect\Yfund})$}\label{su4}

\begin{displaymath}
\begin{array}{|l|c|cccc|} 
                                        \hline
       &  SU(4)       & SU(3) & U(1)_{A} & U(1)_{B} & U(1)_{R}   \\    \hline
A      & \Yasymm    & \Yfund  &    0      &  1      &    0    \\
Q      & \Yfund    &  1       &    1      &  -3      &      0  \\
{\bar Q} & \overline{\Yfund} & 1 &   -1    &  -3    &   0       \\  \hline
Q{\bar Q}  & 1  & 1 & 0& -6 & 0  \\
AA_{sym} & 1 & \Ysymm &   0      & 2 & 0               \\
AAQ{\bar Q}_{anti}& 1 & \Yasymm &  0       & -4 & 0               \\
AAA{\bar Q}{\bar Q}& 1  & 1 & -2 & -6 & 0  \\
AAAQQ & 1  & 1 & 2 & -6 & 0 \\  
\hline
\end{array}
\end{displaymath}
The constraint is: 
\[1/6  (AA_{sym})^3 (Q{\bar Q})^2 + 4 (AA_{sym})(AAQ{\bar Q}_{anti})^2 + 
64 (AAA{\bar Q}{\bar Q}) (AAAQQ) =\Lambda^{8}(Q{\bar Q}) \]

\subsubsection{$SU(5)$ with $2\, \protect\Yasymm +\overline{\protect\Yasymm} +
               \overline{\protect\Yfund}$}\label{su5}

\begin{displaymath}
\begin{array}{|l|c|cccc|}                                            \hline
       &  SU(5)       & SU(2) & U(1)_{A} & U(1)_{B} & U(1)_{R}    \\    \hline
A      & \Yasymm    & \Yfund  & 1       &   0 & 0             \\
{\bar A} & \overline{\Yasymm} & 1 &  -2  & 1 & 0               \\ 
{\bar Q} & \overline{\Yfund} & 1 &   0   & -3 & 0               \\  \hline
A{\bar A} & 1  & \Yfund  &     -1     & 1 & 0               \\
A^{2}{\bar A}^{2} & 1  & \Ysymm & -2  & 2 & 0               \\   
{\bar A}^{2}{\bar Q} & 1  & 1  & -4  & -1 & 0               \\
A^{3}{\bar Q} & 1  & \Yfund  & 3  & -3 & 0               \\
A^{4}{\bar A}{\bar Q} & 1  & \Ysymm & 2   &-2 & 0    \\  
A^{2}{\bar A}{\bar Q}^{2} & 1  & 1  & 0     & -5 & 0    \\ \hline
f_{1} = [(A^{2}{\bar A}^{2}) (A^{4}{\bar A}{\bar Q})]_{flavorsym}& 1& & & &\\
f_{2} = [(A^{4}{\bar A}{\bar Q})^2]_{flavorsym}& 1& & & &\\
f_{3} = [(A^{4}{\bar A}{\bar Q})^2 (A{\bar A})^2]_{flavorsym}& 1& & & &\\
f_{4} = [(A^{2}{\bar A}^{2}) (A{\bar A}) (A^{3}{\bar Q})]_{flavorsym}& 1& & & &\\   
f_{5} = [(A{\bar A}) (A^{3}{\bar Q})]_{flavorsym}& 1& & & &\\   
f_{6} = [(A^{2}{\bar A}^{2}) (A^{3}{\bar Q})^2]_{flavorsym}& 1& & & &\\
f_{7} = [(A^{4}{\bar A}{\bar Q}) (A{\bar A}) (A^{3}{\bar Q})]_{flavorsym}& 1& & & &\\
\hline
\end{array}
\end{displaymath}

The constraint is:  \[(2^{10} f_{1} +2^9 f_{3} +2^7 f_{4}) A^{2}{\bar A}{\bar Q}^{2} + (5 f_{5}^{2} + 2^2 f_{6} - 2^{7} f_{2} -2^6 f_{7}) {\bar A}^{2}{\bar Q} = \Lambda^{8}(A^{2}{\bar A}{\bar Q}^{2})\]

\subsubsection{$ SU(6)$ with $ \protect\Ythreea +\protect\Yasymm + 2\,                       \overline{\protect\Yfund}$}\label{su6}

\begin{displaymath}
\begin{array}{|l|c|cccc|}                                            \hline
       &  SU(6)       & SU(2) & U(1)_{A} & U(1)_{B} & U(1)_{R}       \\    \hline         
B  & \Ythreea & 1 & 1  & 0 &0               \\
A  &\Yasymm  & 1 &   0      &1& 0                \\
{\bar Q}&\overline{\Yfund} & \Yfund & -3  & -2 & 0               \\  \hline
S_{1}=A{\bar Q}^{2}  & 1 & 1 &  -6 & -3 & 0               \\
S_{2}=A^{3}& 1 & 1 &   0& 3 & 0               \\
S_{3}=B^{4}& 1 & 1 &  4 & 0& 0               \\
S_{4}=(B^{4}A^{3})& 1 & 1 &  4 & 3 & 0               \\
(BA^{2}{\bar Q})& 1 &\Yfund & -2 & 0 & 0               \\
(B^{2}A{\bar Q}^{2})_{sym}& 1 &\Ysymm & -4 & -3 & 0               \\ 
S_{5}=(B^{2}A{\bar Q}^{2})_{anti}& 1 & 1 &  -4 & -3 & 0               \\
(B^{3}A^{2}{\bar Q})& 1 &\Yfund &  0 & 0 & 0               \\
S_{6}=(B^{4}A{\bar Q}^{2})_{anti}& 1 & 1& -2  & -3 & 0                \\   \hline
f_{1}=[(BA^{2}{\bar Q})(B^{3}A^{2}{\bar Q})]_{flavorsym}&1&1& & & \\
f_{2}=[(B^{2}A{\bar Q}^{2})_{sym} (B^{3}A^{2}{\bar Q})^{2}]_{flavorsym}&1&1& & & \\
f_{3}=[(B^{2}A{\bar Q}^{2})_{sym} (BA^{2}{\bar Q})^{2}]_{flavorsym}&1&1& & & \\
f_{4}=[(B^{2}A{\bar Q}^{2})_{sym}^{2}]_{flavorsym}&1&1& & & \\
\hline
\end{array}
\end{displaymath}

 The constraint is: 
\begin{eqnarray*}
-12 (6 S_{6} +S_{1} S_{3}) f_{1} + 18 f_{2} -27 S_{3} f_{3} - 648 S_{4} f_{4}  - 16 (18 S_{4} +S_{2} S_{3}) S_{5}^{2} +  \\
48 (12 S_{6}- S_{1} S_{3})S_{4} S_{1} + 96 S_{2} S6^{2} = \Lambda^{12} S_{5}
\end{eqnarray*} 
\subsubsection{$ SU(7)$ with $ \protect\Ythreea +\protect\Yfund + 3\,                       \overline{\protect\Yfund}$}\label{su7}

\begin{displaymath}
\begin{array}{|l|c|cccc|}                                            \hline
       &  SU(7)       & SU(3) & U(1)_{A} & U(1)_{B} & U(1)_{R} \\    \hline
B  & \Ythreea & 1 & 0 & 1 & 0               \\
Q  &\Yfund    & 1 & -3 & -10 & 0               \\
{\bar Q}&\overline{\Yfund} & \Yfund & 1 & 0& 0               \\  \hline
Q{\bar Q} &1 & \Yfund &  -2 & -10 & 0              \\ 
B{\bar Q}^{3} & 1 & 1 & 3 & 1 & 0                \\
B^{3}{\bar Q}^{2} & 1 & \Ysymm &  2 & 3 & 0               \\
B^{3}{\bar Q}^{3}Q_{anti} &  1 & 1 & 0 & -7 & 0               \\
B^{4}Q^{2} & 1 & 1 & -6 & -16 & 0               \\
B^{5}{\bar Q}^{2}Q & 1 & \overline{\Yfund} & -1 & -5 & 0   \\
B^{7} & 1 & 1 & 0 & 7 & 0               \\  \hline
f_{1}=[(B^{3}{\bar Q}^{2})^{3}]_{flavorsym}&1&1& & & \\
f_{2}=[(B^{3}{\bar Q}^{2})^{2} (Q{\bar Q})^{2}]_{flavorsym}&1&1& & & \\
f_{3}=[(B^{5}{\bar Q}^{2}Q) (Q{\bar Q})]_{flavorsym}&1&1& & & \\
f_{4}=[(B^{5}{\bar Q}^{2}Q)^{2} (B^{3}{\bar Q}^{2})]_{flavorsym}&1&1& & & \\
\hline
\end{array}
\end{displaymath}

The constraint is:
\begin{eqnarray*}
7 f_{1} (B^{4}Q^{2}) +6 f_{2} (B^{7}) -288 f_{3} (B^{7}) (B{\bar Q}^{3}) 
+ 1008 f_{4} + 12 (B^{7}) (B^{3}{\bar Q}^{3}Q_{anti})^{2} - \\
72 (B^{4}Q^{2}) (B^{7}) (B{\bar Q}^{3})^2 = 
\Lambda^{14}(B^{3}{\bar Q}^{3}Q_{anti})
\end{eqnarray*} 
\subsection{The Quantum Modified $Sp(N)$ Theories}

\begin{table}[h]                     
\begin{center}
\begin{tabular}{|l|l|l|} \hline
$Sp(2N)$ & $(2N+2)\, \Yfund$ & i-quantum modified \\
$Sp(2N)$ & $\Yasymm +4\, \Yfund $ & i-quantum modified \\
$Sp(2N)$ & $\Ysymm =Adj $ & Coulomb branch \\ \hline 
$Sp(4)$ & $2\, \Yasymm +2\, \Yfund $ & c-quantum modified \\
$Sp(4)$ & $3\, \Yasymm  $ & Coulomb branch \\ \hline 
$Sp(6)$ & $2\, \Yasymm  $ & Coulomb branch \\
$Sp(6)$ & $\Ythreea +3\,\Yfund $ & c-quantum modified \\
 \hline
\end{tabular}
\end{center}
\caption{These are all $Sp$ theories satisfying $\sum_j \mu_j -\mu_G =
0$ and the Witten anomaly condition.  We list the gauge group and the
field content of the theories in the first and second column. In the
third column, we indicate which theories are quantum modified.
The prefix ``i'' indicates an invariant quantum modification 
and the prefix ``c'' a covariant quantum modification.  }
\label{SPtable}
\end{table}

\subsubsection{$Sp(4)$ with $2\, (\protect\Yasymm +\protect\Yfund)$}
\label{sp4}

\begin{displaymath}
\begin{array}{|l|c|cccc|}                                         \hline
       &  Sp(4)       & SU(2)_{A} & SU(2)_{Q} & U(1)_{A} & U(1)_{R} \\\hline
A      & \Yasymm     &  \Yfund  & 1& 1 & 0\\
Q      & \Yfund  & 1 &  \Yfund  & -2 & 0 \\    \hline
Q_{1}Q_{2} &  1  &  1 &  1& -4 & 0                  \\
AA_{sym}  &   1  & \Ysymm &  1& 2 & 0             \\
AQ_{1}Q_{2}  &  1  &\Yfund  & 1& -3 & 0 \\
AAQ_{i}Q_{j} &  1  & 1  & \Ysymm & -2 & 0 \\
\hline
\end{array}
\end{displaymath}

The constraint is: 
\[ (AA_{sym})^2 (Q_{1}Q_{2})^2 -4 ((AA_{sym}) 
(AQ_{1}Q_{2})^2)-16 (AAQ_{i}Q_{j})^2   = \Lambda^{6} (Q_{1}Q_{2})\]

\subsubsection{$Sp(6)$ with $ \protect\Ythreea +3\,\protect\Yfund $}
\label{sp6}

\begin{displaymath}
\begin{array}{|l|c|ccc|}                                         \hline
       &  Sp(6)       & SU(3) & U(1)_{A} &      U(1)_{R}  \\    \hline
B   &\Ythreea &  1      & 3 & 0                              \\
Q      & \Yfund  &\Yfund  &  -5 & 0                           \\    \hline
QQ  &   1  & \Yasymm     &  -10& 0                            \\ 
BQ^{3}&  1  &  1 & -12 & 0                                    \\
B^{2}Q^{2}_{sym}&   1  & \Ysymm &  -4 &  0                    \\
B^{4} &  1  &  1 & 12 &  0                                   \\
B^{3}Q^{3} &  1  &  1 & -6 &  0                              \\  
\hline
\end{array}
\end{displaymath}

The constraint is: 

\begin{eqnarray*}
1728 (B^{3}Q^{3})^{2}-8 (BQ^{3})^{2} (B^{4}) + 12
(B^{2}Q^{2}_{sym})^{3} + 3 (B^{4}) (B^{2}Q^{2}_{sym}) (QQ)^{2} =
\Lambda^{8} (BQ^{3})
\end{eqnarray*}
  
\section{Conclusions}\label{conclusions}

Adding a fundamental and an antifundamental matter multiplet to a
theory with a quantum modified moduli space one obtains a theory
satisfying the index condition $\sum_{i=1}^{n} \mu_{i} - \mu_{G} =
2$. If this theory is s-confining, the algebraic constraint defining
the moduli is invariant under all global symmetries, and the invariant
quantum modified moduli (i-QMM) is given by
$F(\phi_i)=\Lambda^p$. However, if the resulting theory is not
s-confining the constraint of the original theory is only covariant
under global symmetries. This gives a covariant quantum modified
moduli (c-QMM), characterized by $F(\phi_i)=\Lambda^p\phi_k$.

Theories with i-QMM are by now commonplace. Less familiar are theories
with c-QMM. In these theories we believe one must take the lagrange
multiplier enforcing the constraint via a superpotential seriously as
a dynamical degree of freedom. But it follows immediately that the
c-QMM has branches, absent at the classical level, for which global
$U(1)_R$ symmetry is broken. 

\vskip1.2cm
{\it Acknowledgments}
\hfil\break
We are grateful to Ken Intriligator, Erich Poppitz, Witold Skiba
and Martin Schmaltz for many
helpful discussions.  This work is supported by the Department of
Energy under contract DOE-FG03-97ER40506.


\appendix


\section{Gauge Invariant Operators in detail}


\subsection{SU(N) Theories}


\subsubsection{$SU(4)$ with $ 3 \protect\Yasymm +(\protect\Yfund +
               \overline{\protect\Yfund})$}\label{app:su4-3as}

The gauge invariant operators:
\begin{eqnarray}
Q{\bar Q}  & = &  Q_{\alpha}{\bar Q}^{\alpha} \\
AA_{sym}  & = &  1/2 (A_{\alpha \beta}^{i} A_{\gamma \delta}^{j} 
\epsilon^{\alpha \beta \gamma \delta} +A_{\alpha \beta}^{j} 
A_{\gamma \delta}^{i} \epsilon^{\alpha \beta \gamma \delta}) \\
AAQ{\bar Q}_{anti}  & = &  A_{\alpha \beta}^{[i} A_{\gamma \delta}^{j]} 
Q_{\eta}{\bar Q}^{\alpha} \epsilon^{\beta \gamma \delta \eta} 
 \\
AAA{\bar Q}{\bar Q}  & = & 1/6 ({\bar Q}^{\alpha} A_{\alpha \beta}^{i} 
{\bar Q}^{\gamma} A_{\gamma \delta}^{j} A_{\eta \iota}^{k}  
\epsilon^{\beta \delta \eta \iota} \epsilon_{i j k}) \\
AAAQQ  & = &  
1/6 (A_{\alpha \beta}^{i} A_{\gamma \delta}^{j} A_{\eta \iota}^{k} 
Q_{\kappa} Q_{\lambda} \epsilon^{\kappa \delta \eta \iota} 
\epsilon^{\alpha \beta \gamma\lambda }\epsilon_{i j k} )
\end{eqnarray}

The gauge and flavor invariant operators are:
\begin{eqnarray}
(AA_{sym})^3  & = &  (AA_{sym})^{ij} (AA_{sym})^{kl} (AA_{sym})^{mn} \epsilon_{ikm} 
\epsilon_{jln} \\
(AAQ{\bar Q}_{anti})^2 (AA_{sym})  & = &  (AAQ{\bar Q})^{[ij]} (AAQ{\bar
Q})^{[kl]}  (AA)^{mn} \epsilon_{i j m} \epsilon_{k l n}
\end{eqnarray}

\subsubsection{$SU(5)$ with $2\, \protect\Yasymm +\overline{\protect\Yasymm} +
               \overline{\protect\Yfund}$}

The gauge invariant operators are:
\begin{eqnarray}
A{\bar A} & = &  (A_{\alpha \beta} {\bar A}^{\alpha \beta})^{i}\\
A^{2}{\bar A}^{2} & = &  (A_{\alpha \beta} A_{\gamma \delta} 
{\bar A}^{\alpha \gamma} {\bar A}^{\delta \beta} )^{\{ij\}} \\
{\bar A}^{2}{\bar Q}  & = &  {\bar A}^{\alpha \beta} {\bar A}^{\gamma
\delta} 
{\bar Q}^{\lambda} \epsilon_{\alpha \beta \gamma \delta \lambda } \\
A^{3}{\bar Q}  & = &  2/3 (A_{\alpha \beta}^{i} A_{\gamma \delta}^{j} 
A_{ \mu \lambda}^{k} {\bar Q}^{\lambda} 
\epsilon^{\alpha \beta \gamma \delta \mu} \epsilon_{j k}) \\
A^{4}{\bar A}{\bar Q}  & = &  (A_{\alpha \beta}^{\{i} A_{\gamma
\delta}^{j\}} 
{\bar A}^{\alpha \gamma} A_{ \mu \lambda}^{[k} {\bar Q}^{\lambda}  
A_{\nu \tau}^{l]} \epsilon^{\beta \delta \mu \nu \tau })\\
A^{2}{\bar A}{\bar Q}^{2}  & = &  (A_{\alpha \beta} {\bar Q}^{\beta} 
A_{\gamma \delta} {\bar Q}^{\delta} {\bar A}^{\alpha \gamma})^{[ij]} 
\end{eqnarray}

The flavor and gauge invariant operators are:
\begin{eqnarray}
f_1  & = &  (A^{4}{\bar A}{\bar Q})^{ij} (A^{2}{\bar A}^{2})^{kl}
\epsilon_{i k} 
\epsilon_{j k} \\
f_2  & = &  (A^{4}{\bar A}{\bar Q})^{ij} (A^{4}{\bar A}{\bar Q})^{kl} 
\epsilon_{i k} \epsilon_{j k} \\
f_3  & = &  (A^{4}{\bar A}{\bar Q})^{ij} (A{\bar A})^{k} (A{\bar A})^{l} 
\epsilon_{i k} \epsilon_{j k} \\
f_4  & = &  (A^{2}{\bar A}^{2})^{ij} (A{\bar A})^{k} (A^{3}{\bar Q})^{l} 
\epsilon_{i k} \epsilon_{j k} \\
f_5  & = &  (A{\bar A})^{i} (A^{3}{\bar Q})^{j} \epsilon_{i j} \\
f_6  & = &  (A^{2}{\bar A}^{2})^{ij} (A^{3}{\bar Q})^{k} (A^{3}{\bar
Q})^{l} 
\epsilon_{i k} \epsilon_{j k} \\
f_7  & = &  (A^{4}{\bar A}{\bar Q})^{ij} (A{\bar A})^{k} (A^{3}{\bar
Q})^{l} 
\epsilon_{i k} \epsilon_{j k} 
\end{eqnarray}

\subsubsection{$ SU(6)$ with $ \protect\Ythreea +\protect\Yasymm + 2\,
\overline{\protect\Yfund}$}

With

\begin{eqnarray}
 Adj_{\tau}^{\delta}  & = &   B_{\alpha \beta \gamma} B_{\mu \nu \tau} 
\epsilon^{\alpha \beta \gamma \mu \nu \delta} 
\end{eqnarray}

it follows:

\begin{eqnarray}
A{\bar Q}^{2}  & = &  1/2 (A_{\alpha \beta} 
{\bar Q}_{\alpha}^{i} {\bar Q}_{\beta}^{j} - 
A_{\alpha \beta} {\bar Q}_{\alpha}^{j} {\bar Q}_{\beta}^{i}) \\ 
A^{3}  & = &  A_{\alpha \beta} A_{\gamma \delta} 
A_{\nu \tau} \epsilon^{\alpha \beta \gamma \delta \nu \tau} \\ 
B^{4}  & = &  Adj_{\alpha}^{\beta} Adj_{\beta}^{\alpha} \\
(B^{4}A^{3})  & = &  Adj_{\alpha}^{\beta} A_{\beta \gamma} Adj_{\mu}^{\nu}
A_{\nu \rho}  A_{\tau \sigma} \epsilon^{\alpha \gamma \mu \rho \tau \sigma} \\ 
(BA^{2}{\bar Q})  & = &  (B_{\alpha \beta \gamma} {\bar Q}^{\gamma} A_{\mu
\nu}  
A_{\tau \sigma} \epsilon^{\alpha \beta \mu \nu \tau \sigma})^{i} \\       
(B^{2}A{\bar Q}^{2})_{sym}  & = &  (B_{\alpha \beta \gamma} {\bar
Q}^{\gamma} 
B_{\mu \nu \tau} {\bar Q}^{\tau}  A_{\delta \sigma} 
\epsilon^{\alpha \beta \mu \nu \delta \sigma})^{\{ij\}} \\
(B^{2}A{\bar Q}^{2})_{anti}  & = & (B_{\alpha \beta \gamma} A_{\tau
\sigma} 
B_{\mu \nu \tau} {\bar Q}^{\nu} {\bar Q}^{\tau} \epsilon^{\alpha \beta
\gamma \tau \sigma \mu})^{[ij]}  \\
 (B^{3}A^{2}{\bar Q})  & = &  Adj_{\alpha}^{\beta} A_{\beta \gamma} 
B_{\mu \nu \tau} {\bar Q}^{\tau} A_{\delta \sigma} 
\epsilon^{\alpha \gamma \mu \nu \delta \sigma})^{i} \\  
(B^{4}A{\bar Q}^{2})_{anti}  & = &  (Adj_{\alpha}^{\beta} B_{\beta \nu
\tau} 
{\bar Q}^{\nu} {\bar Q}^{\tau} B_{\gamma \delta \eta} A_{\rho \sigma} 
\epsilon^{\alpha \gamma \delta \eta \rho \sigma})^{[ij]} 
\end{eqnarray}
The flavor and gauge invariant operators are:

\begin{eqnarray}
f_1  & = &  (BA^{2}{\bar Q})^{i} (B^{3}A^{2}{\bar Q})^{j} \epsilon_{i j} \\
f_2  & = &  ((B^{2}A{\bar Q}^{2})_{sym})^{ij} (B^{3}A^{2}{\bar Q})^{k} 
(B^{3}A^{2}{\bar Q})^{l} \epsilon_{ik} \epsilon_{jl}\\
f_3  & = &  ((B^{2}A{\bar Q}^{2})_{sym})^{ij} (BA^{2}{\bar Q})^{k} 
(BA^{2}{\bar Q})^{l} \epsilon_{ik} \epsilon_{jl}\\
f_3  & = &  1/2(((B^{2}A{\bar Q}^{2})_{sym})^{ij}((B^{2}
A{\bar Q}^{2})_{sym})^{kl}\epsilon_{ik} \epsilon_{jl})
\end{eqnarray}

\subsubsection{$ SU(7)$ with $ \protect\Ythreea +\protect\Yfund + 3\, 
  \overline{\protect\Yfund}$}
With
\begin{eqnarray}
{\bar B}^{\mu \nu \rho \tau}  & = &  B_{\alpha \beta \gamma} 
\epsilon^{\alpha \beta \gamma \mu \nu \rho \tau} \\
 H^{\alpha}_{\tau \mu \nu}  & = &  {\bar B}^{\alpha \beta \gamma
\delta} 
B_{\gamma \delta \tau} B_{\beta \mu \nu}
\end{eqnarray}

it follows:
\begin{eqnarray}
Q{\bar Q}  & = & (Q_{\alpha} {\bar Q}^{\alpha})^{i}   \\
B{\bar Q}^{3}  & = &  (B_{\alpha \beta \gamma} {\bar Q}^{\alpha} 
{\bar Q}^{\beta} {\bar Q}^{\gamma})^{[ijk]} \\
B^{3}{\bar Q}^{2}  & = &  
({\bar B}^{\mu \nu \rho \tau} B_{\mu \nu  \gamma} 
{\bar Q}^{\gamma} B_{\rho \tau \alpha} {\bar Q}^{\alpha})^{\{ij\}}   \\
B^{3}{\bar Q}^{3}Q_{anti}  & = &  ({\bar B}^{\mu \nu \rho \tau} 
B_{\mu \nu  \gamma} {\bar Q}^{\gamma} B_{\rho \alpha \beta} 
{\bar Q}^{\alpha}{\bar Q}^{\beta} Q_{\tau})^{[ijk]}     \\ 
B^{4}Q^{2}  & = &  {\bar B}^{\mu \nu \rho \tau} Q_{\tau} 
B_{\mu \nu  \gamma} {\bar B}^{\alpha \beta \gamma \delta} Q_{\delta} 
B_{\alpha \beta\rho}   \\
B^{5}{\bar Q}^{2}Q  & = &  ({\bar B}^{\alpha \beta \gamma \delta} 
B_{\alpha \beta \rho} {\bar Q}^{\rho} {\bar B}^{\mu \nu \iota
\tau} B_{\mu \nu \sigma} {\bar Q}^{\sigma} B_{\gamma \delta \iota}
Q_{\tau})^{[ij]} \\
B^{7}  & = &  H^{\alpha}_{\tau \mu \nu}  H^{\beta}_{\rho \sigma \lambda} 
B_{\alpha \beta \gamma} \epsilon^{\gamma \tau \mu \nu \rho \sigma \lambda} 
\end{eqnarray}
The flavor and gauge invariant operators are:
\begin{eqnarray}
f_1  & = &  (B^{3}{\bar Q}^{2})^{ij} (B^{3}{\bar Q}^{2})^{kl} 
(B^{3}{\bar Q}^{2})^{mn} \epsilon_{ikm} \epsilon_{jln} \\
f_2  & = &  (B^{3}{\bar Q}^{2})^{ij} (B^{3}{\bar Q}^{2})^{kl} 
(Q{\bar Q})^{m}(Q{\bar Q})^{n} \epsilon_{ikm} \epsilon_{jln} \\
f_3  & = &  1/4((B^{5}{\bar Q}^{2}Q)^{ij} (Q{\bar Q})^{k} \epsilon_{ijk}) \\
f_4  & = &  1/16((B^{5}{\bar Q}^{2}Q)^{ij} (B^{5}{\bar Q}^{2}Q)^{kl} 
(B^{3}{\bar Q}^{2})^{mn} \epsilon_{ijm} \epsilon_{kln}) 
\end{eqnarray}

\subsection{The $Sp(2N)$ Theories}

We recall that there is an invariant tensor,  
\(J^{\alpha \beta} = (1_{N \times N}
\otimes i\sigma^2)^{\alpha \beta} \).  In the following examples
indices have been freely raised using \(J^{\alpha \beta} \).


\subsubsection{$Sp(4)$ with $2\, (\protect\Yasymm +\protect\Yfund) $}

The gauge invariant operators are:
\begin{eqnarray}
Q_{1}Q_{2}  & = &  Q_{1 \alpha} Q_{2}^{\alpha}  \\
AA_{sym}  & = &  (A_{\alpha \beta} A_{\gamma \delta} 
\epsilon^{\alpha \beta \gamma \delta})^{\{a,b\}}  \\
AQ_{1}Q_{2}  & = &  (A_{\alpha \beta} Q_{1}^{\alpha} Q^{\beta})^{a} \\
AAQ_{i}Q_{j}  & = &  (A_{\alpha \beta} Q^{\beta} A_{\gamma \delta}
Q_{\tau} 
\epsilon^{\alpha \gamma \delta \tau})^{[ab] \{ij\}} 
\end{eqnarray}

The gauge and flavor invariant operators are:
\begin{eqnarray}
(AA_{sym})^2  & = &  1/2 (AA_{sym}^{ab}  AA_{sym}^{cd} \epsilon_{ac}  
\epsilon_{bd}) \\
(AA_{sym})(AQ_{1}Q_{2})^2  & = &  AA_{sym}^{ab} 
(AQ_{1}Q_{2})^{c}(AQ_{1}Q_{2})^{d}\epsilon_{ac}  \epsilon_{bd}\\
(AAQ_{i}Q_{j})^2  & = &  1/2 ((AAQ_{i}Q_{j})^{ab} (AAQ_{i}Q_{j})^{cd} 
\epsilon_{ac}  \epsilon_{bd}) 
\end{eqnarray}

\subsubsection{$Sp(6)$ with $ \protect\Ythreea +3\,\protect\Yfund $}

With 

\begin{eqnarray}
{\bar B}^{\alpha \beta \gamma}  & = &  B_{\mu \nu \tau}
\epsilon^{\alpha 
\beta \gamma \mu \nu \tau} 
\end{eqnarray}

it follows:

\begin{eqnarray}
QQ  & = &   (Q_{\alpha} Q^{\alpha})^{[ij]} \\
BQ^{3}  & = &  (B_{\alpha \beta \gamma} Q^{\alpha} Q^{\beta} 
Q^{\gamma})^{[ijk]} \\
B^{2}Q^{2}_{sym}  & = &  ({\bar B}^{\alpha \beta \gamma} B
_{\alpha \beta \tau} Q^{\tau} Q_{\gamma})^{\{ij\}}\\
B^{4}  & = &  {\bar B}^{\alpha \beta \gamma} {\bar B}^{\mu \nu
\tau} 
B_{\alpha \beta \tau} B_{\mu \nu \gamma} \\
B^{3}Q^{3}  & = &  ({\bar B}^{\alpha \beta \gamma} B_{\alpha \beta
\tau} 
Q^{\tau} B_{\gamma \mu \nu} Q^{\mu} Q^{\nu})^{[ijk]}
\end{eqnarray}

The gauge and flavor invariant operators are:
\begin{eqnarray}
(B^{2}Q^{2}_{sym})^3  & = &  (B^{2}Q^{2}_{sym})^{ij}
(B^{2}Q^{2}_{sym})^{kl} 
(B^{2}Q^{2}_{sym})^{mn} \epsilon_{ikm} \epsilon_{jln}\\
 (B^{2}Q^{2}_{sym})(QQ)^2  & = &  (B^{2}Q^{2}_{sym})^{ij} (QQ)^{kl}
(QQ)^{mn} 
\epsilon_{ikl} \epsilon_{jmn}
\end{eqnarray}
  



\end{document}